\documentclass[epj]{webofc}
\usepackage[utf8]{inputenc}
\usepackage[varg]{txfonts}   
\usepackage{booktabs}
\usepackage{xcolor}
\definecolor{darkred}{rgb}{0.4,0.0,0.0}
\definecolor{darkgreen}{rgb}{0.0,0.4,0.0}
\definecolor{darkblue}{rgb}{0.0,0.0,0.4}
\usepackage[bookmarks,linktocpage,colorlinks,
    linkcolor = darkred,
    urlcolor  = darkblue,
    citecolor = darkgreen]{hyperref}
\usepackage{subfigure}

\wocname{EPJ Web of Conferences}
\woctitle{Lattice2017}
\definecolor{red}{rgb}{0.8,0.0,0.0}
\definecolor{green}{rgb}{0.0,0.6,0.0}
\definecolor{darkblue}{rgb}{0.0,0.1,0.7}
\definecolor{brown}{rgb}{0.6,0.1,0.0}
\definecolor{gray}{rgb}{0.6,0.6,0.6}
\definecolor{darkgreen}{rgb}{0.0, 0.545098, 0.0}
\definecolor{purple}{rgb}{0.5,0.0,0.5}
\definecolor{applegreen}{rgb}{0.55, 0.71, 0.0}
\definecolor{babypink} {rgb}{0.64, 0.44, 0.44}
\definecolor{orange}{rgb}{0.9,0.4,0.0}

\newcommand{\st}[1]{{}}                                 

\newcommand{\rF}{{\ensuremath{F}}\xspace}

\newcommand{\rlF}{{F}}

\usepackage{mathtools}
\DeclarePairedDelimiter{\evdel}{\langle}{\rangle}
\newcommand{\ev}{\evdel}

\usepackage{xspace} 
\newcommand{\Pl}[1]{{\ensuremath{{P^{#1}}}}\xspace}

\newcommand{\PlF}{\Pl{\rlF}}

\begin{document}
%
\selectlanguage{english}
\title{%
Confinement study of an SU(4) gauge theory with fermions in multiple representations
}
\author{%
	\firstname{Venkitesh} \lastname{Ayyar}\inst{1} %
	\thanks{Speaker, \email{venkitesh.ayyar@colorado.edu}} \and
	\firstname{Daniel~C.}  \lastname{Hackett}\inst{1}\and
	\firstname{William~I.}  \lastname{Jay}\inst{1} \and
	\firstname{Ethan~T.}  \lastname{Neil}\inst{1,2}
}
\institute{%
	Department of Physics, University of Colorado, Boulder, Colorado 80309, USA
	\and
	RIKEN-BNL Research Center, Brookhaven National Laboratory, Upton, New York 11973, USA
}
\abstract{%
We discuss the phase diagnostics used in our finite-temperature study of an SU(4) gauge theory with dynamical fermions in both the fundamental and two-index antisymmetric representations.
Beyond the usual Polyakov loop diagnostics of confinement, we employ several Wilson flow phase diagnostics.
The first, what we call the ``flow anisotropy'', is known in the literature: the deconfinement transition introduces anisotropy between the spatial and temporal directions, to which the flow is extremely sensitive.
The second, the ``long flow time Polyakov loop,'' is related but novel.
While we do not claim to fully understand this diagnostic, we have empirically found it to be useful as an unusually sharp diagnostic of phase.
}
\maketitle

\section{Introduction}
\label{sec:intro}
Strongly coupled theories with fermions charged under different representations have been proposed as Beyond Standard Model candidates, specifically in the context of Composite Higgs \cite{GEORGI1984216,DUGAN1985299} and partial compositeness \cite{Kaplan:1991dc}. In this work we explore one such lattice BSM model \cite{Ferretti:2014qta,Ayyar:2017qdf}, an SU(4) gauge theory with fermions in the fundamental (\rF, \textbf{4}, quartet) and two-index antisymmetric ($A_2$, \textbf{6}, sextet) representations.
We present our study of the phase diagram at finite-temperature in a companion Proceedings \cite{other_thermo_proceedings}. In these Proceedings, we would like to present the various confinement phase diagnostics employed in our study. Specifically, we discuss a few Wilson flow phase diagnostics.

The method of Wilson flow was introduced as a smearing operation on the gauge fields in a fictitious ``flow time" \cite{Narayanan:2006rf,Luscher:2009eq}. It is commonly used to set the scale for lattice simulations \cite{Luscher:2010iy}. It has also been used to compute the renormalized coupling in lattice gauge theories \cite{Luscher:2011bx}. Recently, the behavior of certain observables under Wilson flow has been used as a diagnostic to explore the confinement transition \cite{Datta:2015bzm,Datta:2016kea,Wandelt:2016oym}. In this work, we propose a new confinement diagnostic, the Polyakov loop at long Wilson flow times.

\begin{figure}
	\centering
	\includegraphics[width=5in]{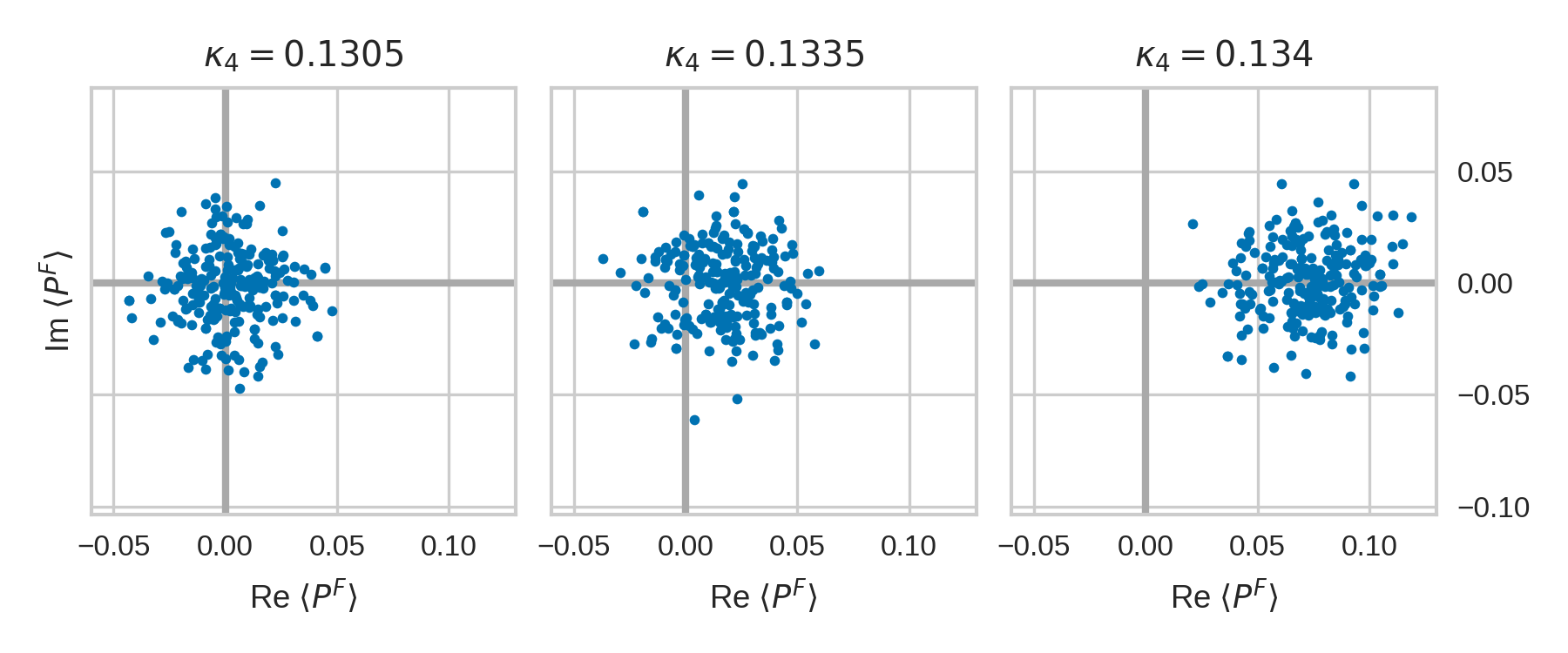}
	\caption{
		Fundamental Polyakov loops in the complex plane for three different $12^3\times6$ ensembles on a slice across the transition varying $\kappa_4$ at constant ${(\beta, \kappa_6)=(7.4,0.1285)}$.
		Each point is the volume-averaged \PlF for one configuration.
		In the left figure, the points are scattered around 0 and $ \langle P^F \rangle = 0 $. This corresponds to a confined ensemble. 
		The right figure where the points are scattered around a central value on the real axis corresponds to a deconfined  ensemble.
		The ensemble shown in the central panel sits almost directly on top of the confinement transition.
	}
	\label{fig:hrpl-scat-ploops}
\end{figure}

We compare the Wilson flow diagnostics with the standard diagnostic for confinement: the (unflowed) Polyakov loop, shown in Fig.~\ref{fig:hrpl-scat-ploops}. Our theory has three bare parameters:  a gauge coupling $ \beta$, and hopping parameters $ \kappa_{4} $ and $\kappa_{6}$ for  the two representations. With fermions in two representations, we can construct Polyakov loops for each representation and look for confinement in each sector separately. In this theory, the confinement transitions for both representations are found to coincide. We discuss this lack of ``phase separation" in more detail in \cite{other_thermo_proceedings}. Here, we focus on diagnosing confinement in the fundamental sector only. 

\section{Flow Anisotropy Diagnostic}
\label{D2}

\begin{figure}
	\centering
	\includegraphics[width=5in]{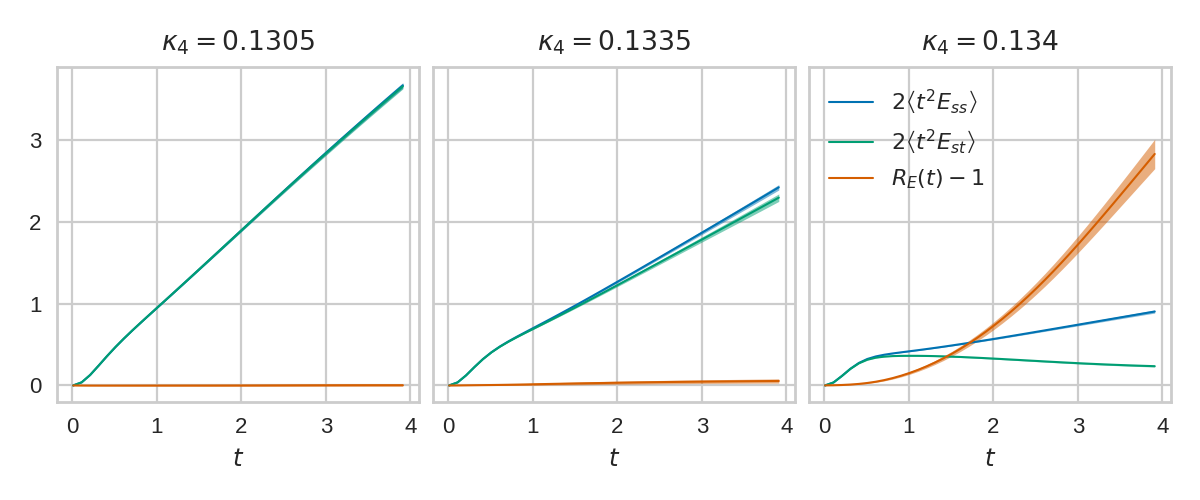}
	\caption{
		Evolution under Wilson flow of the split flow observables $\left< t^2 E_{ss} \right>$ and $\left< t^2 E_{st} \right>$, depicted for three different $12^3\times6$ ensembles on a slice at constant ${(\beta, \kappa_6)=(7.4,0.1285)}$.
		Also depicted is the behavior of $R_E-1$, where $R_E \equiv \left<E_{ss}(t)\right> / \left<E_{st}(t)\right>$.
		The left panel shows typical confined behavior: the split flow observables are degenerate and $R_E=1$, indicating isotropy.
		The central panel shows some hints of impending deconfinement: the split flow observables separate slightly, and $R_E -1 $ becomes nonzero.
		The right panel shows typical deconfined behavior: the split flow observables break apart, and $R_E -1 $ grows rapidly in flow time, indicating anisotropy.
	}
	\label{fig:flow-aniso-obs}
\end{figure}

The flowed observable $\ev{t^2 E(t)}$, where $ E $ represents the energy density, is commonly used to determine the scale for zero-temperature lattices.
When measured on finite temperature lattices, spatial-temporal anisotropy in this same observable can be employed to diagnose the phase \cite{Datta:2015bzm,Datta:2016kea,Wandelt:2016oym}.
Usually, one computes $\ev{t^2 E(t)}$ as a sum over all clover terms at each site, with each clover term's orientation being defined by the two directions it extends in.
This observable can be decomposed as
\begin{equation}
\ev{t^2 E(t)} = \ev{t^2 E_{ss}(t)} + \ev{t^2 E_{st}(t)}
\end{equation}
where $E_{ss} $ is the contribution to $E$ from the three spacelike ($xy$, $xz$, $yz$) clovers and $E_{st}$ is the contribution from the three timelike ($xt$, $yt$, $zt$) clovers.
Previous applications of this diagnostic have used the ``flow splitting observable'',  defined as
\begin{equation}
\Delta(t) = t^2 \langle E_{ss}(t) - E_{st}(t) \rangle .
\label{eqn:flow-split-obs}
\end{equation}
We prefer to look at the ``flow anisotropy observable'', defined as
\begin{equation}
	\langle E_{ss}(t) / E_{st}(t)\rangle  - 1 = \frac{ \Delta(t) }{ \langle t^2 E_{st}(t) \rangle }\equiv R_E(t) -1 \ ,
\end{equation}
which may be better motivated physically, as it relates to the anisotropy in the lattice spacing (see the discussion in Section~\ref{sec:dim-red}). At finite $t$, this quantity provides a sharp diagnostic of phase, as seen in Fig.~\ref{fig:flow-aniso-obs}.
In the confined phase $R_E(t)-1$ is small for all $t$, while in the deconfined phase $R_E(t)-1$ quickly becomes large.
Wilson flow thus appears to amplify anisotropy when applied to deconfined ensembles. This effect is not observed for confined ensembles.

\section{Polyakov Loops at Long Flow Time Diagnostic}
\label{sec:pl-at-long-flow-time}

\begin{figure}
	\centering
	\includegraphics[width=5in]{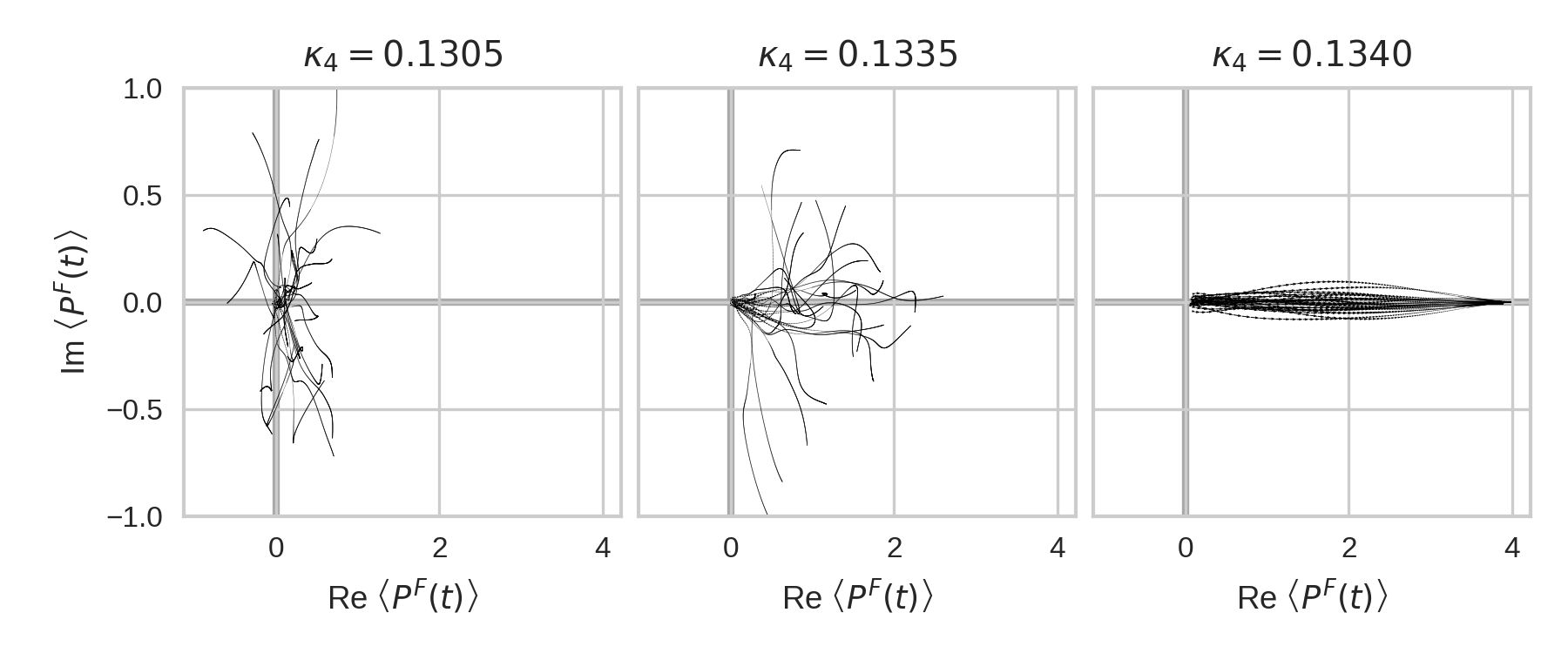}
	\caption{
		Fundamental Polyakov loops under Wilson flow, depicted for three different $12^3\times6$ ensembles on a slice at constant ${(\beta, \kappa_6)=(7.4,0.1285)}$.
		Each panel depicts the complex plane.
		Each line is the evolution of \PlF under Wilson flow on a configuration in the ensemble (i.e., each line is the complex function $\PlF(t)$ where $t$ is the flow time).
		In the two left panels, we see typical confined behavior.
		In the right panel, we see typical deconfined behavior.
		The ensemble shown in the central panel sits almost directly on top of the confinement transition.
	}
	\label{fig:hrpl-squids}
\end{figure}

\begin{figure}
	\centering
	\includegraphics[width=5in]{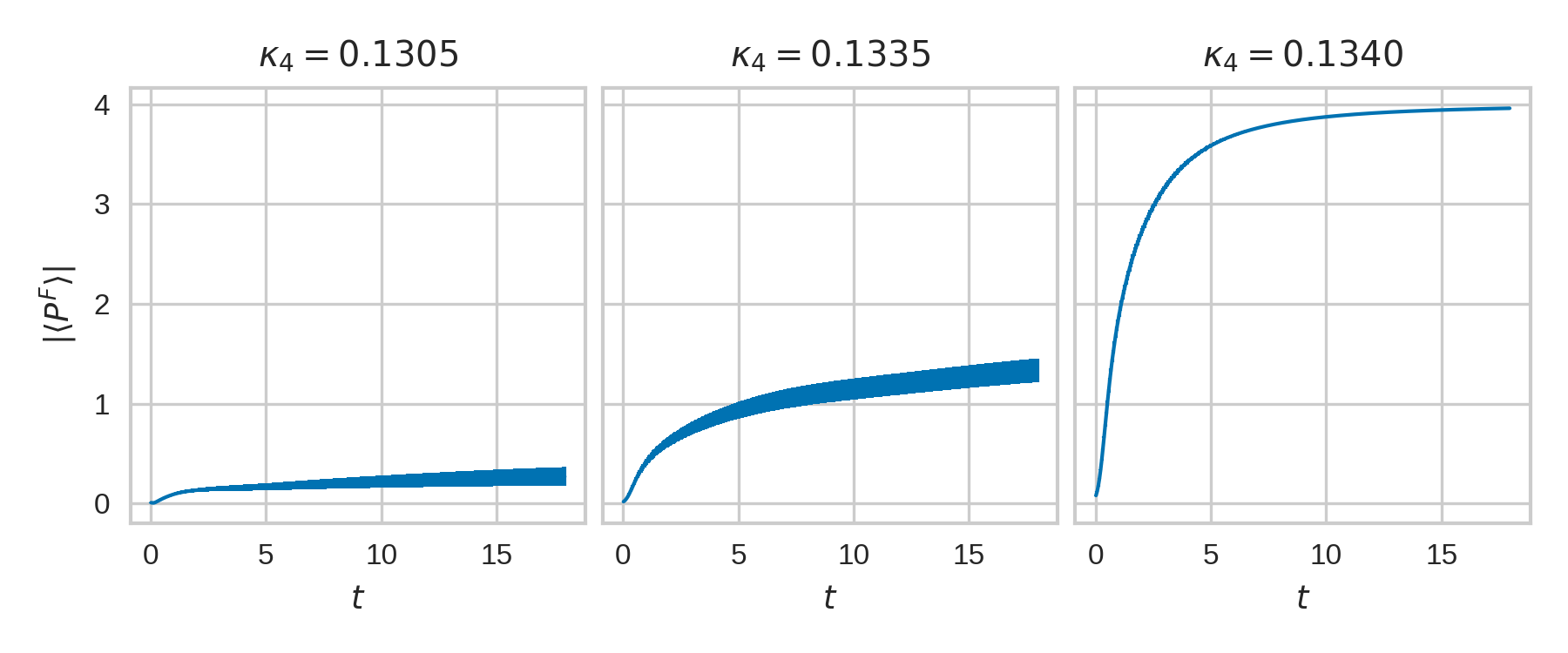}
	\caption{
		Ensemble-averaged evolution of \PlF under Wilson flow, depicted for three different $12^3\times6$ ensembles on a slice at constant ${(\beta, \kappa_6)=(7.4,0.1285)}$.
		In the left panel and center panels, we see typical confined behavior: the Polyakov loop  either does not increase in magnitude or increases very slowly.
		In the right panel, we see typical deconfined behavior: the magnitude of Polyakov loop rapidly approaches its maximum value  ($\max \PlF = d(F)= 4$).
		Despite the $ \kappa_4 $ values for the centre and right panels being very close, the diagnostic does a very good job of distinguishing the phases.
	}
	\label{fig:hrpl-avgs}
\end{figure}

The behavior of the Polyakov loop under smearing has been explored previously \cite{Svetitsky:1997du}.
RG-blocking has been used to sharpen the Polyakov loop signal \cite{Schaich:2012fr}.
More recently, the method of Wilson flow has been applied to the Polyakov loop to remove lattice artifacts and amplify the signal \cite{Datta:2016kea,Schaich:2015psa}.
Further, the method of Wilson flow can be used to obtain renormalized Polyakov loops \cite{Petreczky:2015yta}.
However, all of these studies have stopped short of the extremal ``long flow time case'' (defined below).

First, we define the characteristic flow time ratio as $ c_t = \sqrt{ 8 t } / N_t $ where $ N_t $ is the temporal lattice extent and $ t $ is the Wilson flow time in lattice units.
When $c_t=1$, the smearing radius $\sqrt{8t}$ of the flow is roughly as long as the temporal extent of the lattice.
For the $12^3 \times 6$ lattices shown in all figures in these proceedings, $c_t = 1 $ corresponds to $t \sim 5$.
We define ``long flow time'' as $c_t > 1$. 

For our theory, the behavior of Polyakov loops at long flow time provides an unambiguous diagnostic of the phase.
We observe the following phenomenon, as depicted in Fig.~\ref{fig:hrpl-squids}: when deconfined lattices are flowed, the Polyakov loops order rapidly.
As seen in Fig.~\ref{fig:hrpl-avgs}, this rapid ordering brings the Polyakov loops to their maximal values, $|\ev{\PlF}| \rightarrow d(\rlF) = 4$.
Note that this is the expected behavior of $\PlF$ as the temperature is driven to infinity.
In contrast, $ \langle P^F (t) \rangle $ of confined lattices wander under flow.
The difference in behavior is very sharp. Even ``almost deconfined'' ensembles, like the central panel in Fig.~\ref{fig:hrpl-squids}, will not order at extremely long flow times ($t > 100$).
Increasing the integration resolution of Wilson flow by a factor of ten does not affect this behavior, so it is not an integration artifact.

While we do not understand this exact mechanism, it is easily motivated.
Wilson flow is a smearing operation, so when applied to a finite lattice, we expect it to eventually homogenize the gauge configuration.
Complete homogenization by Wilson flow in the temporal direction ought to take many factors of the characteristic time ratio $c_t$. However, we observe much more rapid temporal homogenization under flow for deconfined ensembles, at $O(1-2)$ times the characteristic flow time. For the deconfined ensemble in  Fig.~\ref{fig:hrpl-avgs}, we see that  $\ev{\PlF}$  approaches its maximal values near $t \sim O(10)$. Meanwhile, this effect is not seen for confined systems.

\section{Possible Mechanism: Dimensional Reduction by Wilson Flow}
\label{sec:dim-red}

\begin{figure}
	\centering
	\includegraphics[width=4in]{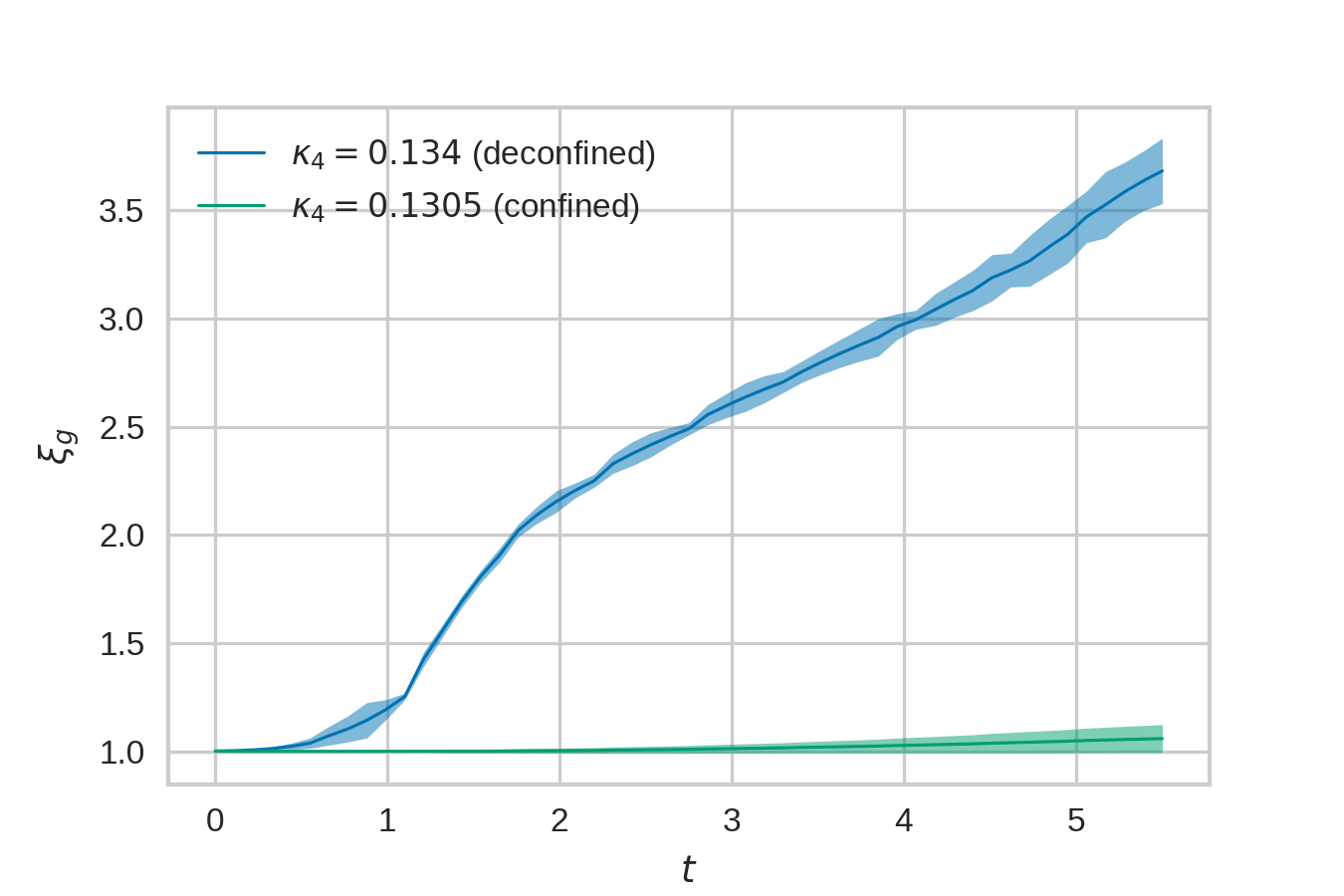}
	\caption{
		This figure shows a plot of  $\xi_g = \frac{a_s}{a_t} $ as a function of flow time $t$. The blue curve represents the deconfined ensemble shown in the right plot of Fig.~\ref{fig:hrpl-avgs}. It grows monotonically and quite rapidly with $ t $. In the long flow time limit, this behavior implies $ a_t \rightarrow 0 $, which suggests movement towards infinite temperature. The green curve represents the confined ensemble shown in the left plot of Fig.~\ref{fig:hrpl-avgs}. In this case, there is no move towards infinite temperature, at least for the range of flow
times we explore.
	}
	\label{fig:xiflow}
\end{figure}

As discussed in Section \ref{D2}, the behavior of the flow anisotropy diagnostic indicates that the process of Wilson flow amplifies anisotropy in the deconfined phase. Meanwhile, as mentioned in Section~\ref{sec:pl-at-long-flow-time}, in the deconfined phase the Polyakov loops behave at long flow time as they would  when the temperature is driven to infinity. In this section, we try to connect these two ideas. We argue that Wilson flow dimensionally reduces deconfined lattices  much more quickly than confined lattices, effectively driving the temperature to infinity.

Using the method described in \cite{Borsanyi:2012zr}, we can use Wilson flow to compute the renormalized anisotropy $\xi_g = a_s/a_t$, where $ a_s $ and $ a_t $ are the lattice spacings in the spatial and temporal directions.
This technique was originally applied to measure the renormalized anisotropy on bare-anisotropic zero-temperature lattices with different bare spatial and temporal couplings ($g_{0s} \ne g_{0t}$). 
Decomposing $\ev{t^2 E(t)}$ in to its spatial and temporal parts, we define   $R_E(t) = \left[ \left<E_{ss}(t)\right> / \left<E_{st}(t)\right> \right] $ as in Section~\ref{D2}. Using anisotropic Wilson flow with parameter $ \xi_{w}$, we can compute $ R_E(t,\xi_w)$ as a function of $ t $ and $ \xi_w $ and obtain $ \xi_{g} $ using the relation $ \frac{R_E(t,\xi_w)}{\xi_w^2}|_{\xi_w=\xi_g}=1$. This gives the relation $ R_E(t) = \xi_g^2 $.

Here, we na\"ively apply the method described above to measure the renormalized anisotropy on bare-isotropic (i.e., $g_{0t} = g_{0s}$) finite-temperature lattices. 
The qualitative behavior of the renormalized anisotropy is easy to read off from the  behaviors of $\ev{t^2 E_{ss}(t)}$ and $\ev{t^2 E_{st}(t)}$ under isotropic flow (as discussed in Section~\ref{D2}).
In the confined phase, $\ev{t^2 E_{ss}(t)}$ and $\ev{t^2 E_{st}(t)}$ remain (approximately) degenerate for all flow times, and so $R_E \approx 1$, and $a_s=a_t$.
However in the deconfined phase, $\ev{t^2 E_{ss}(t)}$ and $\ev{t^2 E_{st}(t)}$ are degenerate at $t=0$ but split apart at finite flow time. Hence $R_E \ne 1$, implying $a_s \ne a_t$.
Observing that $ E_{ss}/ E_{st} \rightarrow \infty $ in the long flow time limit, it follows that $\xi_g \rightarrow \infty$ at long flow times.
This effect is shown in Fig.~\ref{fig:xiflow}. For the confined ensemble, $ \xi_g \sim 1 $, while for the deconfined ensemble, $ \xi_g $ increases rapidly with flow time.
This implies that, under flow, deconfined lattices physically flatten in the temporal direction relative to the spatial direction.  This picture is consistent with a rapid increase in physical temperature.

We do not have a quantitative understanding of why the process of Wilson flow affects confined and deconfined systems differently.
However, all of what we observe follows from the assumption that the Wilson flow amplifies anisotropy in the deconfined phase.

\section{Conclusion}
\label{sec:conclusion}

\begin{figure}[htb!]
	\centering
	\includegraphics[width=5.5in]{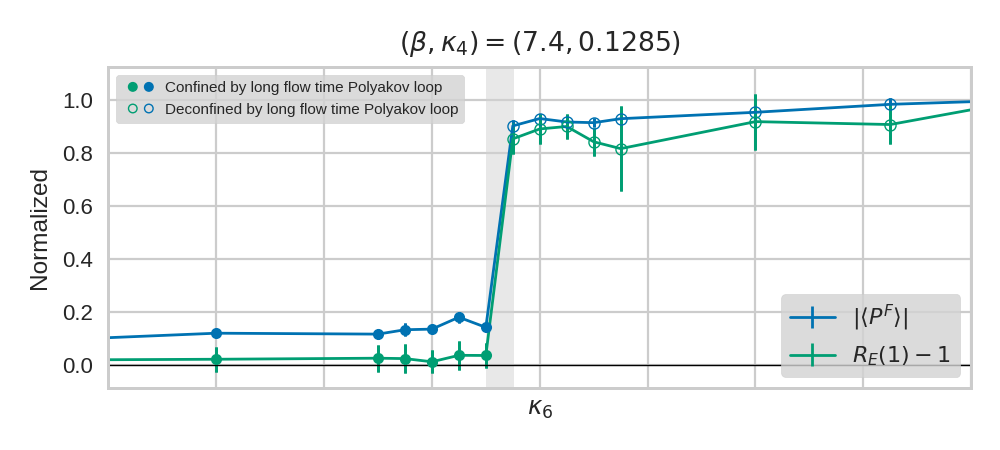}
	\caption{
		This figure compares the three phase diagnostics used in our study, for the fundamental representation. The diagnostics are plotted as a function of the coupling $\kappa_6$ for $ \beta=7.4$ and $ \kappa_4=0.1285$ for a $ 12^3 \times 6 $ lattice. {\bf First:} the blue curve denotes the standard Polyakov loop diagnostic. {\bf Second:} the green curve denotes the ``flow anisotropy" diagnostic. {\bf Third:} the open (closed) circles denote ensembles that are determined to be deconfined (confined) using the ``long flow time Polyakov loop" diagnostic. The shaded region denotes the confinement transition. It is clear that the three diagnostics are consistent.
	}
	\label{fig:diag_comparison}
\end{figure}

We have discussed some Wilson-flow related observables that we use to locate the confinement transition.  As shown in Fig.~\ref{fig:diag_comparison}, all of the diagnostics discussed are consistent with each other.

In particular, the {\it long flow time Polyakov loop} gives a clear and unambiguous (i.e., binary) diagnosis of the phase. We do not yet have a clear understanding of why this observable diagnoses confinement so effectively.
Further analytic work on the effect of flow on finite-temperature systems will be necessary to understand the diagnostic.

It is possible that the first-order thermal transition seen in the multirep lattice theory sharpens the long-flow time diagnostic. For continuous transitions and first-order transitions smoothed to a crossover, this diagnostic may be less effective. We are currently investigating the behavior of this diagnostic in pure gauge theory and in the limiting cases of the multirep theory, where the transition appears to be continuous.

Although further investigation is needed to determine the possible limitations of the long flow time Polyakov loop diagnostic, we believe this could be a convenient tool for future finite temperature studies of lattice gauge theories.

\subsection*{Acknowledgements}
{ 
We would like to thank Thomas DeGrand, Yigal Shamir, and Benjamin Svetitsky for useful discussions. 
This research was supported by U.S.~Department of Energy Grant
Number  under grant DE-SC0010005 (Colorado). Brookhaven National Laboratory is supported
by the U.~S.~Department of Energy under contract DE-SC0012704.
This work utilized the Janus supercomputer, which is supported by the National Science Foundation (award number CNS-0821794) and the University of Colorado Boulder. The Janus supercomputer is a joint effort of the University of Colorado Boulder, the University of Colorado Denver and the National Center for Atmospheric Research.
Additional computations were done on facilities of the USQCD Collaboration at Fermilab,
which are funded by the Office of Science of the U.~S. Department of Energy.
The computer code is based on the publicly available package of the
 MILC collaboration \cite{milc}. 
}

\bibliography{lattice2017}

\end{document}